\documentstyle[12pt]{article}

\newcommand{\rf}[1]{(\ref{eq:#1})}

\begin{document}
\title{  Brane-Worlds and  Cosmology } 
\author{ M. D. Maia\\
Universidade de Bras\'{\i}lia, Instituto  de F\'{\i}sica \\
Bras\'{\i}lia. D.F. 70919-970 \\{maia@fis.unb.br} }

\maketitle

I first met  Mario back in  1965  when we  were  students  at the University of Brasilia, and  I recall   one day  when he  invited me for  a  chat  on the expansion of the  universe.  We never finished that  conversation because  the university was shutdown  soon  after and  we  went to different places. However, Mario always  remained  faithful to  his  cosmological  quest. On his  $60^{th}$ birthday,  I find it rewarding   to have the chance to continue that chat,  in a  much larger  and  complex world  and  under the  brighter light of   type Ia supernovas. I hope we will never end.

\begin{abstract}
The Friedmann equation for  a  FRW-brane-world in a flat bulk is  derived  and  applied to the accelerated expansion of the universe. 
\end{abstract}

\section{Brane-Worlds}
It has been recently noted that   the predicted  scale of  $10^{15}$ TeV  for  quantum gravity is  only  a  conjecture   without  experimental support, and that the only  experimentally  verified   scale  of gauge interactions in four  dimensions lies  within  the TeV scale.  Therefore,  the assumptions that gravitation    becomes  strong at  the TeV scale,  while  the standard gauge interactions remain  confined to the four dimensional space-time,  does  not  conflict with the today's  experimental data  \cite{Arkani,Randall}.   These  ideas, aimed to solve the  the hierarchy problem   of the  fundamental interactions, were originally inspired by  the  Horawa-Witten  M-theory \cite{Witten}.  

We may  identify  four basic principles of the  brane-world program: The first one, of  phenomenological nature,   sets the  fundamental scale  of interaction, including gravity  to the TeV scale. The other three  principles are of  theoretical nature, stating that the space-time (or, the brane-world)  is  embedded in a higher dimensional bulk space; that   ordinary matter  and  the  standard gauge interaction remain confined to the brane-world; and that  the  extra dimensions are probed by gravitons.

Under these  assumptions, the extra  dimensions  can be  large and sometimes  non-compact.  This follows from an  simple argument using the Einstein-Hilbert  action for  the brane-world  $V_{4}$, assuming that the bulk has product topology  $ V_{D}\sim V_{4} \times B_{N}$,  where  $B_{N}$  is the space generated by the extra dimensions 
\[
A=  \int{\cal R}\sqrt{-{\cal G}} dx^{4+N}\approx \frac{1}{M_{*}^{2+N}},\;\;\;
 G_{*}=\hbar c/ M_{*}^{2}, \;\;\;   \hbar=c=1
\] 
where  $M_{*}$ denotes the  equivalent to  Planck's mass in the  higher dimensional space.

On the other hand,  assuming that  ordinary Einstein's gravity remains   confined  to the space-time  $V_{4}$,  we find  that
\[
A= \int{\cal R}\sqrt{-\cal G}d^{4}x  V\approx  \frac{1}{M_{Pl}^{2}}V \;\;\;\;
 G=\hbar c/ M_{Pl}^{2}, \;\;\;   \hbar=c=1
\] 
where $V$ is the volume of  $B_{N}$.
Comparing the two expressions  and denoting by $\ell$ the typical length of the  extra dimensions, we obtain
\[
 \ell  \approx  \frac{M_{Pl}^{2/N}}{M_{*}^{1+2/N}}
\]
If  $M_{*}\approx  1$TeV  e  $M_{Pl}\approx  10^{15}$TeV,  then $\ell   \approx \; 10^{13} cm $ when  $N=1$. This  is  approximately the  diameter of the solar system,  quite unsuitable as an internal dimension.  On the other hand if    $N=2$, a  six dimensional bulk,  we obtain 
the sub millimeter size $ \ell \approx 10^{-2}$cm.  Larger number of dimensions would produce  even smaller  values for $\ell$.  

One immediate consequence is the  modification of Newton's gravitational theory at  small distances. In fact, replacing  $M_{Pl}$ in the Newtonian gravitational potential,  we find
\[
U = \frac{m_{1}m_{2}}{M_{pl}^{2}}\frac{1}{r}=
\frac{m_{1}m_{2}}{M_{\ast}^{2+N}}
\frac{1}{\ell^{N }r}
\]
For  large  values of  $r$  as compared with   $\ell$,  we obtain  the usual Newtonian theory. However, when    $ r  \approx  \ell $  we find
\[
U= \frac{m_{1}m_{2}}{M_{\ast}^{2+N}} \frac{1}{r^{N+1 }}
\]
so that  for  $N=2$  the gravitational field  would decay as  $1/r^{3}$ at   distances of the order of  $10^{-2}$cm.  Interesting enough,   Newton's  theory  has  never  been accurately  verified  at  those small   distances. A number of  experiments  to verify this  are  currently being  assembled \cite{Newmann}.

Other  notable consequences  are the possible generation of  (high frequency)  gravitational  waves  and the   detection of   black-holes (and worm-holes)  in the space-time  at the  TeV scale,   estimated  to  occur sometime around   2006 at the LHC \cite{Giddings:2}.   Cosmology  would also be  a  laboratory to  observe these  phenomena,   where  the presence of  TeV gravitons  at the core of galaxies   could  give a  new  estimate  for the vacuum energy. Even at the classical level,  we may be  already observing  the  effect  of the  extra  dimensions  in the form of  the accelerated expansion of the universe.

\section{Some Geometrical Aspects}
The brane-worlds program  has been implemented mostly on particular models  where the   bulk  has  a fixed  geometry and the  space-time  has a  specific metric ansatz. Nonetheless,  it is possible to  give  a fairly  general description  based on the   four mentioned   principles. Here  we present only a brief sketch \cite{ME}.

Consider that we have  a $D$-dimensional  bulk $V_{D}$ in which we have  a  local and isometric  embedded brane-world $V_{4}$ given by the  embedding map   $Z : V_{4}\rightarrow  V_{D}$  whose derivative map satisfies the  embedding equations
 \begin{equation}
 {\cal Z}^{\mu}_{,i}{\cal Z}^{\nu}_{,j}{\cal G}_{\mu\nu}=g_{ij},\;\;
{\cal Z}^{\mu}_{,i}\eta^{\nu}_{A}{\cal G}_{\mu\nu}=g_{iA},\;\; 
{\eta}^{\mu}_{A}{\eta}^{\nu}_{B}{\cal G}_{\mu\nu}=g_{AB}   \label{eq:embed}
\end{equation}
where  ${\cal G}_{\mu\nu}$ is the metric of  $V_{D}$ in arbitrary coordinates, 
$g_{AB}$ denotes the metric of the space orthogonal to $V_{4}$  and 
 $g_{iA}=s^{M}A_{iAM}$, where 
\begin{equation}
A_{iAB}=\eta^{\mu}_{A,i}\eta^{\nu}_{B} {\cal G}_{\mu\nu}  \label{eq:A}
\end{equation}
The embedding associates an   extrinsic curvature to  $V_{4}$,  defined for each  direction $\eta_{A}$ by
\begin{equation}  
b_{ijA}=-{\cal Z}^{\mu}_{,i}{\eta}^{\nu}_{A;j}{\cal G}_{\mu\nu} \label{eq:KijA}
\end{equation}
Contrarily  to  the Riemannian curvature, the extrinsic curvature  gives  a measure of the  deviation  from the  brane-world and its  tangent plane at any point, which we call  the  {\em bending} of the geometry.  Thus,  we may have a bent  space  even if  it is  flat in the Riemannian sense.   We may define a mean  curvature  for each  normal  direction to  $V_{n}$  by $h_{A}=g^{ij}\kappa_{ijA}$. The total mean curvature is $h^{2} =g^{AB}h_{A}h_{B}$. 

The   integrability conditions  for the embedding equations  \rf{embed}  are the Gauss-Codazzi-Ricci  equations, respectively
 \begin{eqnarray}
R_{ijkl}& =& 2g^{MN}b_{i[kM}b_{jl]N} +{\cal R}_{\mu\nu\rho\sigma}{\cal Z}^{\mu}_{,i}{\cal Z}^{\nu}_{,j}{\cal Z}^{\rho}_{,k}{\cal Z}^{\sigma}_{,l}\nonumber \\
b_{i[jA,k]}& =& g^{MN}A_{[kMA}b_{ij]N} +{\cal R}_{\mu\nu\rho\sigma}
{\cal Z}^{\mu}_{,i} \eta^{nu}{\cal Z}^{\rho}_{,j}{\cal Z}^{\sigma}_{,k}\label{eq:GCR}\\
2A_{[jAB;k]} &+&  2g^{MN}A_{[jMA}A_{k]NB} \nonumber\\
 &+& g^{mn}b_{[jmA}b_{k]nB} +{\cal R}_{\mu\nu\rho\sigma}{\cal
Z}^{\rho}_{,j} {\cal Z}^{\sigma}_{,k}\eta^{\nu}_{A}\eta^{\mu}_{B}= 0 \nonumber
\end{eqnarray}
 From \rf{embed}, we obtain
\begin{equation}
g^{ij}{\cal Z}^{\mu}_{,i}{\cal Z}^{\nu}_{,j}  ={\cal G}^{\mu\nu}
-g^{AB}\eta^{\mu}_{A}\eta^{\nu}_{B}  \label{eq:INV1}
\end{equation}
Applying this, the contractions of  Gauss'  equation gives 
\[
R ={\cal R} - (\omega^{2} + h^{2})  -2 g^{AB}\frac{\partial h_{A}}{\partial s^{B}}
\]
where   we have denoted  $\omega^{2}=b_{ijA}b^{ijA}$  \cite{ME}.
The divergence  term can be discarded  under  a volume integration  in the extra dimensions  and  assuming that the  boundary of the integration region is  minimal  (corresponding to  $h_{A}=0$).  After  adding the  Lagrangian for the  confined matter in the right hand  side,  we obtain the  brane-worlds Lagrangian
 \begin{equation}
{\cal L}(g)= R\sqrt{g} =  8\pi G{\cal  L}_{m}\sqrt{g} +{\cal R}\sqrt{{\cal G}}+ (k^{2} + h^{2}) \sqrt{{g}}\label{eq:EH}
\end{equation}
The corresponding Einstein's  equations  for the brane-world can be calculated directly from 
the  contractions  of  Gauss'  equations
\begin{equation}
R_{ij} -\frac{1}{2}R g_{ij}  = 8\pi G T_{ij}^{m} + ({\cal  R}_{\nu\rho}-\frac{1}{2}{\cal R}{\cal  G}_{\nu\rho})Z^{\nu}_{,i}Z^{\rho}_{,j}  +Q_{ij} +S_{ij  } \label{eq:Einstein}
\end{equation}
where we have denoted  the extrinsic term  by
\begin{equation}
Q_{ij} =g^{MN}(b^{m}{}_{iM}b_{j m N} -h_{M}b_{ij N})  -\frac{1}{2}(\omega^{2}- h^{2})g_{ij}
\end{equation}
and  an  additional term which also depends  on  the Riemann and  Ricci curvatures of the bulk
\begin{equation}
S_{ij}  =g^{AB}{\cal R}_{\mu\nu} \eta^{\mu}_{A}\eta^{\nu}_{B}g_{ij} +g^{AB}{\cal R}_{\mu\nu\rho\sigma}\eta^{\mu}_{A}\eta^{\sigma}_{B}Z^{\nu}_{,i}Z^{\rho}_{,j}
-\frac{1}{2} g^{AB}g^{MN}{\cal R }_{\mu\nu\rho\sigma} \eta^{\mu}_{A}\eta^{\sigma}_{B}\eta^{\nu}_{M}\eta^{\rho}_{N} g_{ij}
\end{equation}

Now, the  confinement hypothesis implies that  ordinary matter,  is  confined to the brane-world. Therefore,  if the bulk is  a solution of  the high dimensional Einstein's  equations,  it  is  either  a  vacuum  or  cosmological constant solution
\[
({\cal  R}_{\mu\nu}-\frac{1}{2}{\cal R}{\cal  G}_{\mu\nu})Z^{\mu}_{,i}Z^{\nu}_{,j} = \Lambda g_{ij}
\]
Placing  this  on  the left hand side of  \rf{Einstein}  we obtain
\begin{equation}
R_{ij} -\frac{1}{2}R_{ij}-\Lambda g_{ij}  = 8\pi GT^{m}_{ij} +Q_{ij}+S_{ij}  \label{eq:bw}
\end{equation}
To these equations we add the  conservation law,  obtained from the contracted Bianchi identity
\begin{equation}
8\pi G T^{ij}{}_{;j}   +  S^{ij}{}_{;j}=0,  \;\;\ Q^{ij}{}_{;j}=0  \label{eq:CONS}
\end{equation}
 We  focus  our  attention to    $Q_{ij}$  which  depends  essentially on the  extrinsic  curvature and  does not depend  on the Riemannian  curvature.    

This is  as far as we can go without  being  more specific. A full account of   the  brane-worlds equations in a  general  bulk, including  the confinement and  some quantum perspectives can be found on  \cite{ME} and references therein.

\section{FRW as a Brane-World  in Flat Bulk}
The standard FRW  cosmological model is   known to  be   embeddable in a flat  5-dimensional  flat bulk  with metric signature  $(4,1)$  and  $g_{55}=1$  \cite{Rosen}.  
We use the following parametrization
\[
dS^{2}=g_{ij}dx^{i}dx^{j}=-dt^{2}  +a^{2}[dr^{2}+f(r)(d\theta^{2} +sen^{2}\theta d\varphi^{2})]
\]
where  $f(r)=sen r, r, senh r$   corresponding to   $k=1,0,-1$ respectively. The confined matter is the perfect fluid  $T_{ij}=(p+\rho)U_{i}U_{j}  +pg_{ij}$,  where  $U_{i}=\delta_{i}^{0}$  in commoving coordinates.
The   embedding equations  \rf{GCR}  become  simply
\begin{eqnarray}
R_{ijkl}  &= &2b_{i[k}b_{l]j}\label{eq:G}\\b_{i[j;k]}& =& 0 \label{eq:C}
\end{eqnarray}
Codazzi's  equation may be  easily  solved, with    solution \cite{MR}
\[
b_{ab}=\frac{b}{a^{2}}g_{ab}, \;\;  \mbox{and}  \;\; b_{44}= \frac{-1}{\dot{a}}\frac{d}{dt}(\frac{b}{a})
\]
or,  after denoting  $B=\dot{b}/b$ e  $H=\dot{a}/a$,  it follows that
\[
\omega^{2}=b^{ij}b_{ij} = \frac{b^{2}}{a^{4}} ( \frac{B^{2}}{H^{2}}-2\frac{B}{H}+4)), \;\; h=g^{ij} b_{ij} =\frac{b}{a^{2}}(\frac{B}{H}+2)
\]
Therefore
\[
\omega^{2} - h^2  = -\frac{6b^{2}}{a^{4}}\frac  BH
\]
and 
\[
Q_{aa} =\frac{b^{2}}{a^{4}}\left( \frac{2B}{H}-1\right)g_{aa},\;\;Q_{44}= \frac{3b^{2}}{a^{4}}\;\;\mbox{e}\;\; Q=tr(Q_{ij})=\frac{6b^{2}}{a^{4}}(\frac{B}{H}-1)
\]
In this example  we have    $S_{ij}=0$  and  the dynamical equations  \rf{Einstein} for the FRW metric  become
\begin{eqnarray*}
\frac{\ddot{a}}{a}=-\frac{4\pi G}{3}(3p+\rho) +\frac{1}{3}(Q_{44}+\frac{1}{2}Q)\\
a\ddot{a}+2\dot{a}^{2} +2k=4\pi G(p-\rho) -(Q_{ab}-\frac{1}{2}Qg_{ab})  
\end{eqnarray*} 
After eliminating $\ddot a$,   we obtain  the modified Friedmann's  equation
\begin{equation}
\label{eq:fri}\dot a^2+k=\frac{8\pi G}{3}\rho  a^{2}+ \frac{b^2}{a^2} \label{eq:Friedmann}
\end{equation}
Notice that  $b(t)= b_{11}$, measures  how the universe bends from the tangent space  along
the radial direction.  To this  we add  the conservation law  and the state  equation $p= (\gamma-1)\rho$,    which are combined in
\[
\frac{\dot{\rho}}{\rho}  +  3\gamma\frac{\dot{a}}{a} -\frac{3}{\rho}\frac{d}{dt}(\frac{b^{2}}{a^{4}})=0
\]
It is possible to  add  an extra  equation relating  $b(t)$ to  the matter density. This  follows from  the   sandwich conjecture in general relativity using  junction conditions  such as Israel's  adapted to higher dimensions in the Randall-Sundrum moldels. However,  these  junction  conditions are not unique and  their  use lead to instabilities \cite{Deruelle}. The application of these conditions to  homogeneous and isotropic  cosmologies  produces a  modified  Friedmann  equation,  with an  added a term proportional to the square of the density \cite{Binetruy}.

On the other hand,  the    accelerated expansion  of the universe  may be  seen  as  an  evidence for  equation \rf{Friedmann}, suggesting that    the universe  bends  in proportion to its  expansion:
\[
b(t) = k'  a(t) 
\]
In  fact,   we   easily see that the negative  constant $ k'^{2} $ on the left  side of \rf{Friedmann}  acts  as  an  accelerator  for the  expansion, even  when  $k=0$.

 Another  interesting comment  is that  the   signature of the  bulk may vary as a  response to the quantum   fluctuations \cite{ME}. This  means that the  term  $k'^{2}$ in \rf{Friedmann}  may   change sign   with the consequent  shift  from accelerated expansion to contraction.


\begin{thebibliography}{10}

\bibitem{Arkani} N. Arkani-Hamed et al, Phys. Lett. {\bf \underline{B429}, 263
 (1998)},\\ Phys. Rev. Lett {\bf \underline{84}, 586 (2000)}
\bibitem{Randall} L. Randall \&  R. Sundrum, Phys. Rev. Lett.
{\bf\underline{83}, 3370 (1999)},\\ Phys. Rev. Lett.
{\bf\underline{83}, 4690 (1999)}
\bibitem{Witten} P.  Horava  \& E. Witten,  Nucl. Phys.  \underline{B475}, 94 (1996).
\bibitem{Newmann} R. Newmann,  in  Matters of Gravity, hep-th/0002021
\bibitem{Giddings:2} S. Giddings,   hep-ph/0110127
\bibitem{ME}  M. D. Maia \&  Edmundo  M.  Monte,  hep-th/0110088
\bibitem{Rosen} J. Rosen,  Rev.  Mod. Phys.   \underline{37}, 204 (1965)
\bibitem{MR}  M. D Maia  \&  W. L. Roque Phys. Lett.   \underline{A139}, 121 (1989)
\bibitem{Deruelle}  N. Deruelle  et al,  hep-th/0010215
\bibitem{Binetruy} P. Binetruy el  al,  hep-th/9905012
\end{thebibliography}
\end{document}